\documentclass[particles,article,accept,pdftex,moreauthors]{Definitions/mdpi} 
\usepackage{amssymb,color,graphicx,bm,mathrsfs,color,amsmath,slashed,comment}

\newcommand{\bce}{\begin{center}} \newcommand{\ece}{\end{center}}
\newcommand{\be}{\begin{equation}} \newcommand{\ee}{\end{equation}}
\newcommand{\bea}{\begin{eqnarray}}
\newcommand{\eea}{\end{eqnarray}}

\definecolor{red}{rgb}{0.8,0,0}
\definecolor{violet}{rgb}{0.4,0,0.4}
\definecolor{green}{rgb}{0,0.5,0.0}
\definecolor{navy}{rgb}{0.0,0.0,0.6}
\definecolor{orange}{rgb}{0.8,0.2,0.0}

\usepackage[normalem]{ulem}  

\firstpage{1} 
\pubvolume{1}
\issuenum{1}
\articlenumber{0}
\pubyear{2026}
\copyrightyear{2026}
\externaleditor{Firstname Lastname} 
\datereceived{26 March 2026} 
\daterevised{15 May 2026} 
\dateaccepted{22 May 2026} 
\datepublished{ } 
\pdfoutput=1

\Title{Isentropic Hybrid Stars in the Nambu--Jona-Lasinio Model: Effects of Neutrino Trapping}

\Author{{Andrea Sabatucci} 
 $^{1,}$*\orcidA{} and  Armen Sedrakian $^{1,2}$\orcidB{}}

\address{%
$^{1}$ \quad Institute of Theoretical Physics,  University of Wroc\l{}aw, Maxa-Borna 9,
  50-204 Wroc\l{}aw, Poland; {armen.sedrakian@uwr.edu.pl} 
\\
$^{2}$ \quad Frankfurt Institute for Advanced Studies,
Ruth-Moufang-Stra\ss e, 1, 60438 Frankfurt am Main,
Germany}

\corres{Correspondence: andrea.sabatucci@uwr.edu.pl}

\abstract{Binary neutron star mergers and proto-neutron stars provide
  unique environments where dense matter is hot, lepton-rich, and
  potentially undergoes a transition from hadronic to deconfined quark
  matter. We investigate the thermodynamics and stellar properties of
  hybrid matter under such conditions. The hadronic phase is described
  within a covariant density functional framework, while the quark
  phase is modeled using a Nambu–Jona-Lasinio (NJL) model that
  includes repulsive vector interactions, the axial $U_A(1)$-breaking
  ’t Hooft determinant interaction, and two-flavor
  color-superconducting (2SC) pairing. The phase transition between
  hadronic and quark matter is constructed using a mixed-phase
  prescription that enforces baryon and lepton number conservation,
  allowing us to follow thermodynamic trajectories at fixed entropy
  per baryon and a fixed lepton fraction. We analyze the phase structure
  of dense matter at a finite temperature and study the composition of
  the hadronic, mixed, and quark phases in both neutrino-trapped and
  neutrino-free regimes. Our results show that neutrino trapping
  significantly modifies the particle composition and shifts the onset
  of deconfinement to higher densities. The mixed phase exhibits a
  density-dependent pressure due to the presence of multiple conserved
  charges. Using the resulting equations of state, we compute static
  stellar configurations and examine the influence of the temperature and
  lepton content on the mass–radius relation in hybrid stars. Hot,
  neutrino-rich configurations are found to have larger radii and
  slightly higher maximum masses than their cold counterparts. As the
  star cools and deleptonizes, its radius contracts at an approximately
  constant baryonic mass, potentially triggering changes in the
  internal phase structure. These results highlight the roles of color
  superconductivity, lepton trapping, and thermal effects in shaping
  the structure and evolution of hybrid stars in transient
  astrophysical environments.}


\begin{document}

\section{Introduction} 
\label{sec:intro}

The advent of multimessenger astronomy, particularly following the
detection of gravitational waves from the binary neutron star merger
GW170817 by the LIGO Scientific Collaboration and the Virgo
Collaboration, has opened up a new window into the physics of matter at extreme densities. During~the late inspiral and merger of binary
neutron stars, the densities and temperatures in the remnant can
significantly exceed those found in cold isolated neutron stars,
potentially triggering a phase transition from hadronic matter to
deconfined quark matter. Such a transition could influence the
post-merger dynamics, the~emitted gravitational-wave spectrum, and~the
lifetime of the hypermassive remnant, making gravitational-wave
observations a powerful probe of the equation of state at
supranuclear densities. In~these environments, the matter is hot and
lepton-rich, with~neutrinos temporarily trapped on dynamical
timescales---conditions similar to those found in core-collapse
supernovae and in the early evolution of proto-neutron stars. The~presence of trapped neutrinos modifies the composition and
thermodynamic properties of dense matter, affecting the threshold for
deconfinement and the structure of the resulting hybrid
configurations. Consequently, understanding the behavior of quark
matter at a finite temperature and lepton fraction is essential for
describing the evolution of merger remnants, proto-neutron stars, and~supernova cores, as~well as for interpreting future gravitational-wave and neutrino observations, which may reveal signatures of the
hadron–quark phase transition in astrophysical~environments.

The aim of this work is to investigate the equation of state and
composition of \mbox{three-flavor} quark matter under thermodynamic conditions
relevant for proto-neutron stars and the early stages of binary neutron star merger remnants,
where temperatures are high and~neutrinos remain trapped in the dense
matter. To~describe the quark phase, we employ a
vector interaction-enhanced Nambu--Jona-Lasinio (NJL) model that includes
both the ’t Hooft determinant interaction, accounting for axial $U(1)$
symmetry breaking, and~pairing in the two-flavor color-superconducting (2SC)
channel involving two colors and two flavors. This model was
previously studied, e.g., in ref.~\cite{Bonanno2012}, in~the
zero-temperature limit appropriate for cold neutron stars assuming a
strong first-order phase transition from hadronic to quark
matter. Here, we generalize this approach to the finite entropy case by~constructing a thermodynamically consistent transition between
hadronic and quark matter, implementing a mixed-phase prescription in which each phase satisfies local electric charge neutrality. The~existence of such a phase in proto-neutron stars, where global lepton number conservation in the presence of trapped neutrinos permits the formation of a mixed phase even for large values of the surface tension at the hadron–quark interface, was first proposed in refs.~\cite{Hempel_Pagliara,Pagliara_PRL}. This approach allows us to explore the properties of hybrid matter in
equilibrium at a finite temperature and lepton fraction. Using this
methodology, we demonstrate the existence of stable isothermal stellar
configurations and analyze their internal compositions in the presence
of neutrino trapping, highlighting the impacts of trapped leptons on
the onset of deconfinement and the structure of hybrid~stars.

Several studies have investigated the thermodynamics and astrophysical
implications of the hadron–quark phase transition at a finite
temperature and entropy under conditions similar to those in our study. For~instance, ref.~\cite{Carlomagno2024,Carlomagno_Blaschke} analyzed hybrid stars with color-superconducting quark cores using a nonlocal
chiral quark model combined with the DD2 relativistic density functional
for hadronic matter. Their isentropic stellar sequences reveal that
the presence and melting of diquark condensates strongly influence the
thermal behavior across the deconfinement transition and may lead to
the appearance of disconnected hybrid branches, giving rise to the
“thermal twin’’ phenomenon. There are complementary approaches that
attempt to provide unified descriptions of dense matter. In~this
context, ref.~\cite{Celi2025} introduced an equation of state that
incorporates hadronic and quark degrees of freedom within a single
relativistic mean-field framework and dynamically models the
deconfinement transition via a Polyakov loop-inspired field, enabling
consistent studies of the QCD phase diagram and proto-neutron star
evolution under astrophysical constraints. Thermal effects on the
baryon–quark phase transition have also been examined in
ref.~\cite{Ghaemmaghami2023}, which combined a statistical mean-field
hadronic model with the NJL model for quark matter. Their results
highlight the important role of neutrino trapping, which can soften
the equation of state in the coexistence region and modify the
thermodynamic conditions for the onset of deconfinement in hot hybrid
stars. Other works explore the QCD phase structure and stellar
composition across different evolutionary stages. Using a nonlocal
three-flavor Polyakov–NJL model, ref.~\cite{Malfatti2019}
investigated the possible existence of spinodal instabilities and the
location of the critical end point and~studied proto-neutron star
matter at a finite temperature and lepton fraction, finding that hot
stellar configurations may contain hyperons and $\Delta$-isobars, while
deconfined quarks appear only in cold neutron stars. Finally,
ref.~\cite{Mariani2017} examined the hadron–quark transition in hybrid
stars across different evolutionary stages using a field correlator
method description of quark matter. Their analysis shows that hybrid
stars satisfying the two-solar-mass constraint can form and that the
deconfinement transition may occur during the later cooling stages of
proto-neutron stars.  We will discuss the specific differences between
our study and the works mentioned above in more detail in the
Conclusions section.  Recently, ref.~\cite{Gholami2025,RGNJL_stars} used a
renormalization group-motivated three-flavor color-superconducting NJL
model to study the structure of cold hybrid stars. While their
analysis focuses on zero-temperature configurations and is therefore
not directly applicable to the hot, neutrino-trapped conditions
considered here, it demonstrates that cutoff artifacts present in
standard NJL models at high densities can be mitigated within
refined~formulations.

Nevertheless, it is already evident at this stage that there is a strong demand for further studies of the interplay between temperature,
composition, and~the phase structure of dense matter in compact stars---in~particular, in~the rarely studied case of trapped neutrino phases. This
is of particular interest also regarding dissipation in binary neutron star mergers with a trapped neutrino component; see refs.~\cite{Alford:2021qmc,Most:2021}
for trapping in the hadronic component. The quark phases have been studied in
the untrapped case; see~refs.~\cite{Hernandez:2024bulkvisc,Alford2024PhRvDLett} and
the references therein. These works also underscore the importance of employing
consistent microphysical models of strongly interacting matter in
order to reliably interpret current and future multimessenger
observations of neutron stars, including those from gravitational-wave
signals, electromagnetic counterparts, and~neutrino~emissions.

This paper is organized as follows. In~Section~\ref{sec:baryonic_eos}, we
briefly describe the equation of state employed for the low-density
baryonic phase. The~high-density quark phase in the 2SC state is discussed in
Section~\ref{sec:quark_eos}. The~treatment of the hadron–quark phase
transition in the presence of neutrino trapping and the construction
of the mixed phase are presented in
Section~\ref{sec:PT_trapped_neutrinos}. In~Section~\ref{sec:results}, we
present our numerical results for the thermodynamics, composition,
phase transition, and~stellar structure under the conditions
considered in this work. Finally, our conclusions and a comparison
with (some) previous studies are given in Section~\ref{sec:discussion}.

\section{Baryonic~Matter} 
\label{sec:baryonic_eos}
The confined phase of hadronic matter, which occupies the
lower-density region of the stellar interior, is described within the
framework of covariant density functional theory. In~this phase, we
consider matter composed of neutrons, protons, electrons, and, when
relevant, trapped neutrinos in $\beta$-equilibrium. For~the hadronic
equation of state, we adopt the results of ref.~\cite{Tsiopelas2024EPJA}, where
a suite of finite-temperature EoS tables was constructed for
applications in numerical simulations of compact-object environments
such as core-collapse supernovae, proto-neutron stars, and~binary
neutron star mergers. These tables are based on a covariant density
functional model that incorporates the full $J^P=1/2^+$ baryon octet
and is consistent with current nuclear physics and astrophysical
constraints. {In this work,} 
 we will use a subset of the tables that includes only nucleonic degrees of freedom.

In this framework, the~baryon–meson coupling constants are density-dependent and are chosen such that the slope of the symmetry energy is
$L_{\rm sym}=50\,\mathrm{MeV}$ and the skewness parameter is
$Q_{\rm sat}=400\,\mathrm{MeV}$, corresponding to the DDLS family of
parametrizations~\cite{Li_2023}. The~model is constructed to describe
matter over a wide range of densities, temperatures, and~electron
fractions relevant to astrophysical simulations. At~sub-saturation
densities, the~uniform matter EoS is smoothly matched to a model
describing inhomogeneous nuclear matter. The~resulting tables provide
thermodynamically consistent information about the composition and
thermodynamic properties of dense matter as a function of the density,
temperature, and~lepton fraction. The~model employed in this work  is available
through the CompOSE database under the name TSO
DDLS(50)-N~\cite{Tsiopelas2024EPJA}. This EoS has been previously
applied in numerical simulations of binary neutron star mergers~\cite{Gieg2025} and
provides a realistic description of the hadronic phase in the density
regime preceding the onset of quark~deconfinement.

The EoS with trapped neutrinos, for~different values of the neutrino chemical potential 
$\mu_{\nu_e}$, is constructed using the CompOSE tables by enforcing $\beta$-equilibrium and subsequently including the neutrino contribution in the thermodynamic quantities, such as pressure, entropy, and~energy density. In~practice, $\beta$-equilibrium is implemented by imposing \mbox{the condition}
\bea
\mu_{Le}(T, \varrho, Y_Q) = \mu_{\nu_e},
\eea
where $\mu_{Le}$, provided by the CompOSE tables, is a function of the baryon density 
$\varrho$, temperature $T$, and~baryonic charge fraction $Y_Q$.

\section{2SC Quark~Matter}
\label{sec:quark_eos}

To describe 2SC quark matter, we
employ an extended local NJL model including vector
interactions. Neglecting electromagnetic effects, the~effective
Lagrangian is
\begin{eqnarray}
\label{eq:NJL-Lagrangian}
\mathcal{L}_{\text{NJL}} &=&
\bar{\psi}(i\gamma^\mu\partial_\mu-\hat m)\psi
+ G_S \sum_{a=0}^8 \left[(\bar{\psi}\lambda_a\psi)^2
+(\bar{\psi} i\gamma_5 \lambda_a\psi)^2\right] \nonumber\\
&+& G_V(\bar{\psi} i\gamma^\mu\psi)^2
+ G_D \sum_{\gamma,c}
\left[\bar{\psi}^a_\alpha i\gamma_5
\epsilon^{\alpha\beta\gamma}\epsilon_{abc}(\psi_C)^b_\beta\right]
\left[(\bar{\psi}_C)^r_\rho i\gamma_5
\epsilon^{\rho\sigma\gamma}\epsilon_{rsc}\psi^s_\sigma\right] \nonumber\\
&-& K \Bigl\{
\det_f[\bar{\psi}(1+\gamma_5)\psi]
\det_f[\bar{\psi}(1-\gamma_5)\psi]
\Bigr\},
\end{eqnarray}
where $\psi_\alpha^a$ denotes quark fields with flavor indices
$\alpha=u,d,s$ and color indices $a=r,g,b$, and~$\hat m=\mathrm{diag}_f(m_u,m_d,m_s)$
is the current quark mass matrix. Besides~the standard scalar–pseudoscalar interaction ($G_S$), the~model includes a repulsive vector interaction ($G_V$), a~diquark pairing channel ($G_D$), and~the ’t Hooft determinant term ($K$), which breaks the axial $U_A(1)$ symmetry.
The parameters are chosen as $m_{u,d}=5.5~\text{MeV}$,
$m_s=140.7~\text{MeV}$, $\Lambda=602.3~\text{MeV}$,
$G_S\Lambda^2=1.835$, and~$K\Lambda^5=12.36$~\cite{RehbergPhysRevC}.
The diquark coupling is taken as $G_D=G_S$, while the vector coupling is generally varied in the range $\eta_V = G_V/G_S=(0.8-1.2)$, but,~in this work, we show results only
for the $\eta_V = 1$ case.  The~vacuum pressure is
$P_0=4263.8\;\text{MeV}/\text{fm}^3$. { The parameters are 
fixed by fitting to empirical meson and quark properties within the SU(3) NJL framework. The~fitted observables include the vacuum masses and decay constants of the pion, kaon, and~eta mesons, as~well as the mass of the eta-prime meson—the latter being especially sensitive to the $U_A(1)$ anomaly encoded in the 't~Hooft determinant term.} An additional bag constant $B^*$ is introduced to fix the deconfinement transition density; for the case of $\eta_V=1$, we adopt $B^*=-20\;\text{MeV}/\text{fm}^3$, corresponding to $\varrho_T\simeq3.2\varrho_0$ at zero temperature. {This particular value puts the critical point above which the 
first-order phase transition vanishes on the 50~MeV isotherm.}

At intermediate baryon densities, the~energetically favored phase is the 2SC phase, where up and down quarks form Cooper pairs in a color antisymmetric $\bar{\mathbf{3}}$ channel while strange quarks remain unpaired~\cite{Alford2008RvMP}. The~corresponding pairing gap is
\bea
\Delta_c \propto G_D
\left\langle(\bar{\psi}_C)^a_\alpha i\gamma_5
\epsilon^{\alpha\beta c}\epsilon_{abc}\psi^b_\beta\right\rangle .
\eea

Chiral symmetry breaking is characterized by the condensates
$\sigma_\alpha\propto\langle\bar{\psi}_\alpha\psi_\alpha\rangle$,
which generate constituent quark masses
\bea
M_\alpha = m_\alpha - 4G_S\sigma_\alpha +2K\sigma_\beta\sigma_\gamma,
\eea
with $(\alpha,\beta,\gamma)$ cyclic in the flavor space.
Vector interactions produce mean fields $\omega_0$ and $\phi_0$
that shift the quark chemical potentials,
\bea
\mu_{f,c}=\frac{1}{3}\mu_B+\mu_Q Q_f+\mu_3 T_3^c+\mu_8 T_8^c,
\eea
where $\mu_Q$, $\mu_3$, and~$\mu_8$ enforce electric and color~neutrality.

The thermodynamic potential is evaluated in the mean-field
approximation (see, e.g., refs.~\cite{Alford2008RvMP,Ruster2005PhRvD,Blaschke2005PhRvD,Gomez2006PhRvD,Bonanno2012}),
and the condensates and chemical potentials are obtained from its
stationarity conditions. This procedure leads to 
the pressure of quark matter, which reads
\bea\label{eq:Pressure}
P &=& \frac{1}{2\pi^2}\sum_{i=1}^{18}\int_0^\Lambda dk\,k^2
\left[|\epsilon_i|
+2T\ln\!\left(1+e^{-|\epsilon_i|/T}\right)\right]
+4K\sigma_u\sigma_d\sigma_s \nonumber\\
&-& \frac{1}{4G_D}\sum_{c=1}^3|\Delta_c|^2
-2G_S\sum_{\alpha=1}^3\sigma_\alpha^2
+\frac{1}{4G_V}(2\omega_0^2+\phi_0^2)
+\sum_{l=e^-,\mu^-}P_l - P_0 - B^*,
\eea
where $\epsilon_i$ denotes quasiparticle energies, $P_l$ is the lepton
pressure, and~$B^*$ controls the onset of~deconfinement.

In this work, we focus on conditions relevant for neutron-star transients such as binary neutron star mergers and core-collapse supernovae, {considering entropies per baryon $S=1$–$2$ (in units of $k_B$) and fixed lepton fractions corresponding to neutrino-trapped matter; see refs.~\cite{Sumiyoshi2007,Fischer2009,Foglizzo2015,OConnor2018,Burrows2020,Pascal2022,Faber2012,Rosswog2015,Baiotti2017,Baiotti2019,Blacker2024} for comparison. We also present results at zero temperature, which provide a reference for cold neutron star configurations.}
 
\section{Phase Transition with Trapped~Neutrinos}
\label{sec:PT_trapped_neutrinos}

In hot neutron star matter, the neutrino mean free path becomes smaller
than the stellar radius once the temperature reaches the order of a
few MeV~\cite{Hofmann2025}. Neutrinos are therefore trapped in the
core and participate in $\beta$-equilibrium, modifying the composition
of dense matter. Under~these conditions, the~phase transition between hadronic and quark matter is governed by the conservation of the baryon number and electron lepton number. First-order phase transitions involving two globally conserved charges,  first discussed by Glendenning in ref.~\cite{Glendenning},  require generally a Gibbs~construction.

We assume that electric charge neutrality is enforced locally in each phase. This assumption is appropriate when the surface tension at the hadron–quark interface is sufficiently large~\cite{Alford:2006bx,Palhares2010,Lugones:2013ema}. Reliable calculations of the surface tension remain challenging and have so far been performed only within schematic models; see, for~example, the studies of the interface between chirally broken nuclear matter and an approximately chirally symmetric phase (interpreted as quark matter) in refs.~\cite{Schmitt_1,Schmitt_2}. 
Under the assumption of local charge neutrality, the~system is characterized by two globally conserved quantities: the baryon number and electron lepton number~\cite{Hempel_Pagliara}. 

At a fixed temperature, the~thermodynamic state of each phase is therefore specified by the pressure, $P(T,\mu_B,\mu_{Le})$.
For the hadronic phase, $\beta$-equilibrium and charge neutrality~\mbox{imply}
\begin{equation}
\mu_n=\mu_p+\mu_e-\mu_{\nu_e}, \qquad Y_p=Y_e .
\end{equation}
{Introducing} 
 the chemical potentials associated with the conserved
charges $\mu_B$, $\mu_Q$, and~$\mu_{Le}$, the~particle chemical
potentials can be written as
\begin{equation}
\begin{split}
\mu_n &= \mu_B, \\
\mu_p &= \mu_B+\mu_Q, \\
\mu_e &= -\mu_Q+\mu_{Le}, \\
\mu_{\nu_e} &= \mu_{Le}.
\end{split}
\end{equation}
In the quark phase, the chemical potentials of the quarks are
\begin{equation}
\mu_u=\frac{1}{3}\mu_B+\frac{2}{3}\mu_Q, \qquad
\mu_d=\mu_s=\frac{1}{3}\mu_B-\frac{1}{3}\mu_Q .
\end{equation}
The phase transition occurs when thermal, chemical, and~mechanical
equilibria between the two phases are satisfied,
\begin{equation}
T^I=T^{II}, \qquad
\mu_B^I=\mu_B^{II}, \qquad
\mu_{Le}^I=\mu_{Le}^{II}, \qquad
P^I(T,\mu_B,\mu_{Le})=P^{II}(T,\mu_B,\mu_{Le}).
\end{equation}
These conditions define a phase boundary in the space
$(P,\mu_B,\mu_{Le})$ separating the two~phases.

Astrophysically relevant configurations correspond to matter with a fixed electron lepton fraction $Y_{Le}$. Typical values are $Y_{Le}\sim0.4$ in proto-neutron stars and $Y_{Le}\sim0.1$ in post-merger remnants. {These values follow from numerical simulations; see refs.~\cite{Sumiyoshi2007,Fischer2009,Foglizzo2015,OConnor2018,Burrows2020,Pascal2022,Mezzacappa2023} 
for the supernova context and refs.~\cite{Faber2012,Rosswog2015,Baiotti2017,Baiotti2019,Blacker2024} for the binary neutron star merger context. }
Fixing $Y_{Le}$ determines a relation between
$\mu_B$ and $\mu_{Le}$ and allows the construction of the equation of
state along this trajectory.  Within~the mixed phase, the pressure
follows the phase boundary, while the baryon density is given by
\begin{equation}
\bar{\varrho}_B=(1-\chi)\varrho_B^I+\chi\varrho_B^{II},
\end{equation}
where $\chi$ is the volume fraction of quark matter. The~latter is
determined from the lepton fraction constraint
\begin{equation}
Y_{Le}=\frac{(1-\chi)\varrho_{Le}^{I}+\chi\varrho_{Le}^{II}}
{\bar{\varrho}_B}.
\end{equation}
Once $\chi$ is known, the~thermodynamic quantities in the mixed phase, such as the energy density $\epsilon$ and the entropy per volume $s$, follow from
\begin{equation}
\bar{\epsilon}=(1-\chi)\epsilon^I+\chi\epsilon^{II}, \qquad
\bar{s}=(1-\chi)s^I+\chi s^{II}.
\end{equation}
In contrast to the Maxwell construction, the~pressure varies across
the coexistence region because two conserved charges are~present.

Finite-temperature neutron stars and merger remnants are commonly
described by isentropic equations of state, characterized by
fixed entropy per baryon. To~construct such EoSs in the presence of
trapped neutrinos, we first compute a set of isothermal EoSs using the
Gibbs construction described above. The~EoS at fixed entropy is then
obtained by interpolating between these isothermal~results.

{The procedure used to obtain isentropic trajectories is the following. We first generate 25 thermodynamically consistent isothermal EoSs with temperatures ranging from 2 to \mbox{50 MeV} using the Gibbs construction described above. For~each fixed temperature, the~coexistence line in the $(\mu_{B},\mu_{Le})$ plane is determined from the Gibbs equilibrium conditions. The~collection of all boundary points defines the phase boundaries in the $(T,\mu_{B},\mu_{Le})$ space.
Isentropic trajectories are then constructed separately for the hadronic and quark phases by imposing fixed entropy per baryon $S=\bar{S}$, with~$\bar{S}=1$-$2$. Their intersections with the mixed-phase surface determine the entrance and exit points of the mixed phase along a given~isentrope.

Inside the mixed phase, the~isentropic EoS is obtained by interpolating between neighboring isothermal mixed-phase EoSs. In~practice, for~each point along the isentropic trajectory, we identify the isothermal mixed-phase curves with temperatures lying between the two boundary temperatures and determine the thermodynamic state satisfying the constraint $S=\bar{S}$. Since the interpolation is performed between thermodynamically consistent Gibbs-constructed EoSs and the entropy constraint is imposed explicitly at each point, the~resulting isentropic trajectories remain smooth and thermodynamically consistent across the mixed phase.}

For comparison, we also analyze the case without trapped neutrinos. In~this limit, the phase transition reduces to the standard Maxwell
construction with local charge neutrality. While the pressure remains
constant throughout the coexistence region for isothermal matter, in~isentropic matter, both the temperature and the pressure vary with
the density, allowing the mixed phase to extend over a finite region
inside the~star.

To compute stellar properties, each equation of state is complemented
at low densities by matching it to an isentropic counterpart in the
HS(DD2) model described in ref.~\cite{HEMPEL2010210}. This model is
based on the improved nuclear statistical equilibrium treatment of
nucleons and nuclear clusters and is widely used in simulations of hot
and dense astrophysical matter. The~matching is performed at baryon
densities $\varrho_B < 0.5\varrho_0$, corresponding to temperatures of
approximately $5\,\mathrm{MeV}$ and $10\,\mathrm{MeV}$ for the $S=1$ and
$S=2$ EoSs, respectively~\cite{Sumiyoshi2007,Fischer2009,Foglizzo2015,OConnor2018,Burrows2020,Pascal2022,Mezzacappa2023,Faber2012,Rosswog2015,Baiotti2017,Baiotti2019,Blacker2024}. In~this low-density regime, neutrinos are not
included. Although~this matching is not strictly thermodynamically
consistent, it is not expected to significantly affect the stellar
properties derived in this work.  Finally, the~macroscopic stellar
observables are obtained by integrating the Tolman–Oppenheimer–Volkoff
equations~\cite{Tolman,OppVolkoff}.

In this study, we neglect the contribution of the muon lepton sector. In~proto-neutron star matter, a vanishing muon lepton fraction, $Y_{L_{\mu}=0}$, is usually imposed. This condition does not strictly prohibit the presence of muons, but~it implies that any finite muon density must be compensated by an excess of muon antineutrinos. Muons can therefore appear once the electron chemical potential becomes sufficiently large for weak processes involving the muon sector to be energetically allowed. In~typical proto-neutron star conditions, however, the~muon population is expected to remain subdominant. Moreover, the~consistent inclusion of the muon sector in a phase transition analysis under Gibbs conditions introduces an additional conserved charge and~hence an extra chemical potential associated with the muon lepton number. This significantly increases the complexity of the construction. Given the expected minor role of muons, and in order to keep the analysis tractable, we neglect their contribution and focus instead on the impact of trapped electron neutrinos on the nature of the phase transition. The~effect of the muon sector can be systematically investigated in future~work.


\section{Results}
\label{sec:results}

\begin{figure}[H]
    \includegraphics[width=0.75\linewidth]{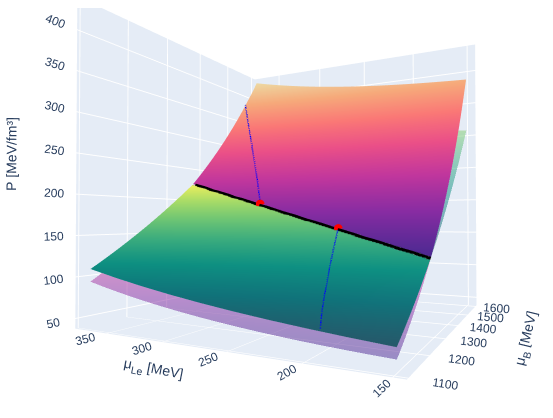}
    \caption{{A 3D plot} 
 of the pressure as a function of the baryon and
      lepton chemical potentials at fixed temperature $T=50$~MeV. The~surfaces
      correspond to the hadronic {(green)} and quark {(purple)} equations of state. Their
      intersection (black solid line) defines the phase boundary
      between the two phases. The~blue dot-dashed line indicates the
      trajectory corresponding to a fixed lepton fraction
      $Y_{Le}=0.4$, while the red circles mark the endpoints of the
      mixed-phase region.}
    \label{fig:3dplot}
\end{figure}

We begin by discussing the thermodynamic properties of the individual
phases and the structure of the mixed phase. Figure~\ref{fig:3dplot}
shows the pressure as a function of the baryon and lepton chemical
potentials at a fixed temperature. The~two surfaces correspond to the
hadronic and quark equations of state. Their intersection defines the
phase boundary where the conditions of thermal, chemical, and~mechanical equilibria between the two phases are simultaneously
satisfied. Along this line, the pressures of the two phases are equal,
while the baryon and lepton chemical potentials take the same values
in both~phases.

\vspace{-2pt}

For astrophysical applications, the~physically relevant thermodynamic
trajectories correspond to matter at a fixed electron lepton fraction
$Y_{Le}$. Such trajectories appear as curves on the pressure surface
in the $\mu_B-\mu_{Le}$ plane. The~intersection of a fixed-$Y_{Le}$
trajectory with the phase boundary determines the onset and end
points of the mixed phase. Between~these two points, the system
undergoes a first-order phase transition, where the two phases coexist
and the thermodynamic quantities are determined by the appropriate
volume fractions of hadronic and quark matter. This construction
provides the basis for determining the equation of state used in the
stellar structure~calculations.

\begin{figure}[H]
\begin{adjustwidth}{-\extralength}{0cm}
\centering
    \subfloat{\includegraphics[scale=0.5]{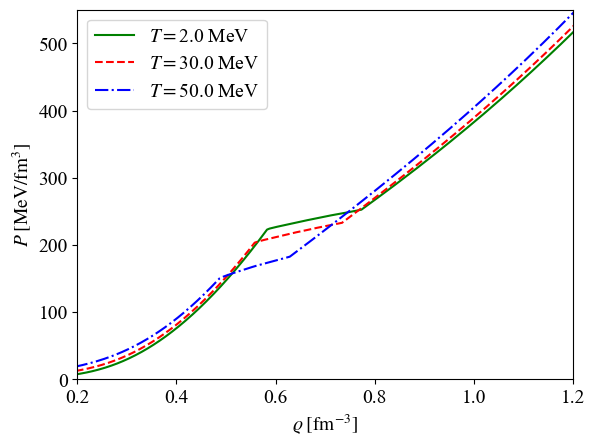}}
     \subfloat{\includegraphics[scale=0.5]{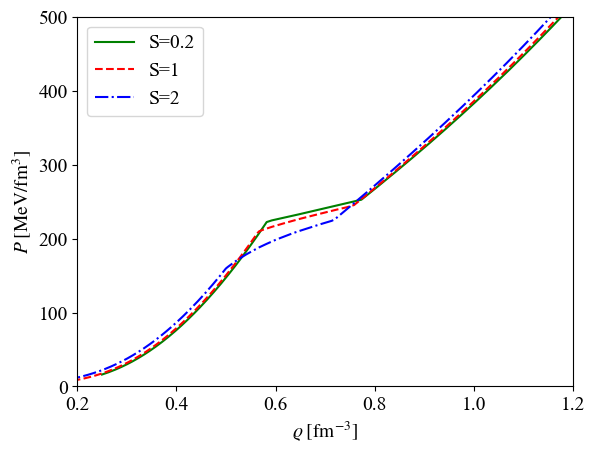}} 
\end{adjustwidth}  
    \caption{The pressure as a function of the baryon density for  a fixed
      lepton fraction $Y_{Le}=0.4$ and different temperatures (\textbf{left panel}) and entropies (\textbf{right panel}) at~fixed vector coupling $\eta_V=1$. }
    \label{fig:2dplot}
\end{figure}

The two-dimensional counterpart of Figure~\ref{fig:3dplot} is displayed in the left panel of Figure~\ref{fig:2dplot}, where the pressure is shown as a function of the baryon density for different temperatures. In~contrast to the standard Maxwell construction, the~pressure does not remain constant throughout the coexistence region but increases
smoothly across the mixed phase. This reflects the fact that the phase
transition is governed by two globally conserved quantities, the baryon
number and lepton number, and~therefore requires a Gibbs construction.
The right panel of Figure~\ref{fig:2dplot} shows the isentropic curves for three representative values of the entropy per baryon.
In the pure hadronic and quark phases, the pressure increases with
the entropy, as~expected from the growing thermal contribution to the
equation of state. Inside the mixed phase, however, the~ordering of
the isentropic curves can be partially reversed. This behavior can be
understood from the thermodynamics of phase equilibrium. The~transition occurs when the Gibbs free energy per baryon of the two
phases is equal. Since the entropy per baryon is the temperature
derivative of the Gibbs free energy, a~phase with larger entropy
decreases its Gibbs free energy more rapidly as the temperature
increases. Because~quark matter typically has larger entropy per
baryon than hadronic matter, increasing the entropy shifts the
coexistence conditions and modifies the density interval over which
the two phases coexist. As~a result, the~pressure in the mixed phase
is determined not only by the thermal stiffness of the individual
phases but~also by the entropy dependence of the phase boundary
itself. This leads to the nontrivial behavior seen in
Figure~\ref{fig:2dplot}.

\vspace{-12pt}

\begin{figure}[H]
\begin{adjustwidth}{-\extralength}{0cm}
\centering
    \includegraphics[scale=0.51]{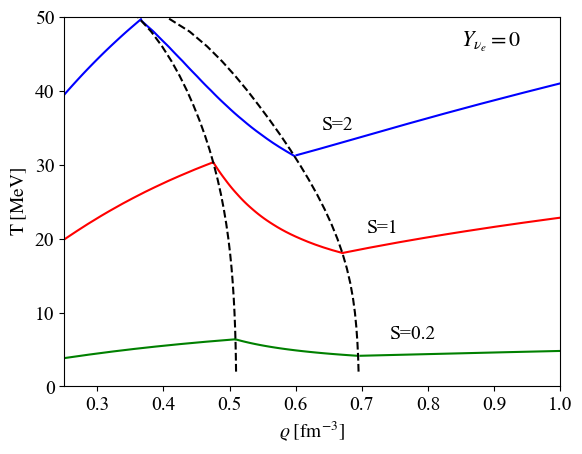}
    \includegraphics[scale=0.51]{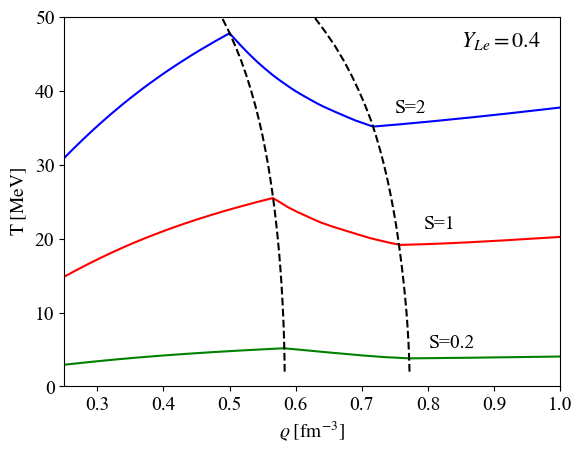}
\end{adjustwidth}
    \caption{{Phase} 
 diagram in the $T$–$\varrho_B$ plane for different
      neutrino content. Several isentropic trajectories are
      shown. The~presence of trapped neutrinos shifts the onset of the
      hadron–quark phase transition to higher baryon densities and
      leads to the appearance of an extended mixed-phase region. {Phase boundaries are shown by black dashed lines.}
    }\label{fig:PD_GV10}
  \end{figure}

  In Figure~\ref{fig:PD_GV10}, we show the constant-entropy trajectories
  in the temperature–density plane, which effectively provide a phase
  diagram for different values of the entropy per baryon $S$. In~the
  baryonic phase, the temperature increases monotonically with the density,
  and, for a given density, it is larger for higher entropy, as~expected
  from the thermodynamic relation between temperature and entropy in
  dense matter.  When the system enters the mixed phase, the~  temperature decreases and eventually aligns with the temperature
  corresponding to the quark phase. This behavior reflects the
  thermodynamics of the phase transition: the onset of deconfinement
  introduces additional degrees of freedom, which increases the
  entropy of the system. At~fixed entropy per baryon, this additional
  entropy must be compensated for by a reduction in the temperature. As~a
  result, the~isentropic trajectories exhibit a drop in temperature
  across the coexistence region. The~effect is more pronounced for
  larger entropy because the thermal contribution to the equation of
  state is stronger, leading to the larger redistribution of entropy
  between the two phases.  Once the system reaches the pure quark
  phase, the~temperature again increases with the density along the
  isentropic trajectory. Thus, the temperature drop observed in the
  coexistence region can be interpreted as a manifestation of the
  entropy difference between the baryonic and quark phases, with~the
  latter typically possessing a larger number of accessible
  microstates. 

\vspace{-2pt}
  
  A comparison between the cases with and without trapped neutrinos in
Figure~\ref{fig:PD_GV10} reveals a noticeable shift in the phase
transition. In~the neutrino-trapped regime, the onset of deconfinement
occurs at higher densities. Physically, this behavior arises because
the presence of trapped neutrinos increases the electron and proton
fractions through the constraint of the fixed lepton number, making the
hadronic phase more stable against the appearance of quark matter. As~a consequence, the~transition is delayed, and~the density range of the
mixed phase is modified compared to the neutrino-free~case.

\begin{figure}[H]
\begin{adjustwidth}{-\extralength}{0cm}
\centering
   \subfloat{\includegraphics[scale=0.55]{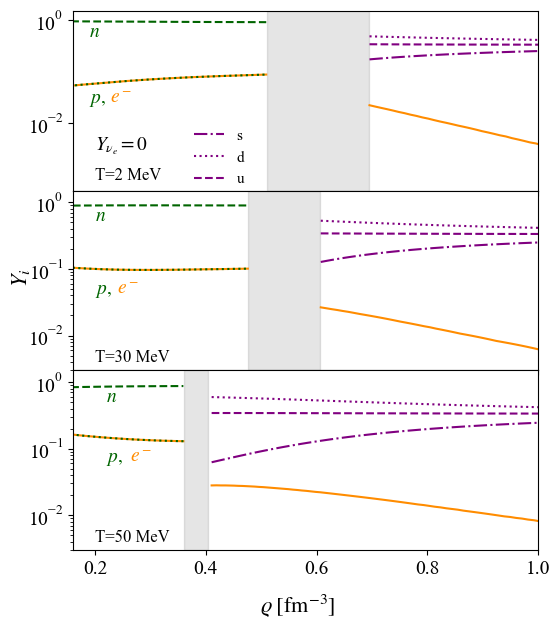}}
    \subfloat{\includegraphics[scale=0.55]{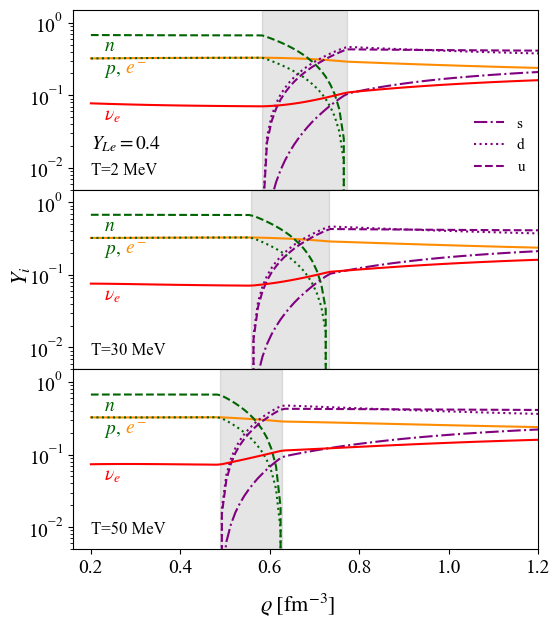}}    
\end{adjustwidth}   
    \caption{{Comparison} 
 of the composition of matter without neutrinos
    (\textbf{left panel}) and with neutrinos at a fixed lepton fraction $Y_{Le}=0.4$ (\textbf{right panel})
      and for different values of the temperature. }
    \label{fig:composition_T}
\end{figure}

  In Figure~\ref{fig:composition_T}, we show the compositions of the baryonic
  and quark phases as a function of the baryon density for three different
  temperatures. The~left panels correspond to matter in
  $\beta$-equilibrium without trapped neutrinos, while the right
  panels show the case with neutrino trapping at a fixed lepton
  fraction $Y_{Le}=0.4$. The~mixed-phase region is indicated by the
  shaded~areas.

 \vspace{-11pt}  

  In the baryonic phase, the electron fraction, and~consequently the
  proton fraction through charge neutrality, is significantly larger
  in the neutrino-trapped case than in the neutrino-free case, where
  standard $\beta$-equilibrium is imposed. In~the trapped regime, the
  neutrino fraction remains approximately constant at $\sim 0.1$,
  remaining subdominant compared with electrons. The~presence of
  trapped neutrinos therefore increases the overall lepton content of
  the matter and modifies the proton–neutron balance in the baryonic
  phase.

  The main effect of the temperature is to shift the onset of the mixed
  phase to lower densities as the temperature increases. At~the same
  time, the~density interval over which the mixed phase exists becomes
  narrower. This behavior is consistent with the thermodynamics of the
  phase boundary discussed above: an increasing temperature favors the
  phase with larger entropy per baryon, which, in this case, is quark
  matter. As~a result, the~transition occurs earlier, while the
  coexistence region becomes~compressed.

  In the neutrino-free case, the~increasing abundance of strange
  quarks at high densities provides a sufficient negative charge to
  neutralize the positive charge of the $u$-quarks, which leads to the
  strong suppression of the electron fraction. In~contrast, when
  neutrinos are trapped, the lepton fraction is fixed, forcing
  electrons to remain present. The~additional negative charge carried
  by electrons is then compensated for by a larger fraction of positively
  charged $u$-quarks.

  Overall, the~presence of neutrinos significantly modifies the charge
  balance of quark matter. The~additional electrons required by the
  fixed lepton fraction increase the negative charge of the system,
  which is compensated for by a larger abundance of positively charged $u$
  quarks. A~similar effect occurs in the baryonic phase, where the
  fixed lepton fraction leads to a substantially larger population of
  electrons and, through charge neutrality, an~equal increase in the
  proton~fraction.

\begin{figure}[H]
    \includegraphics[scale=0.65]{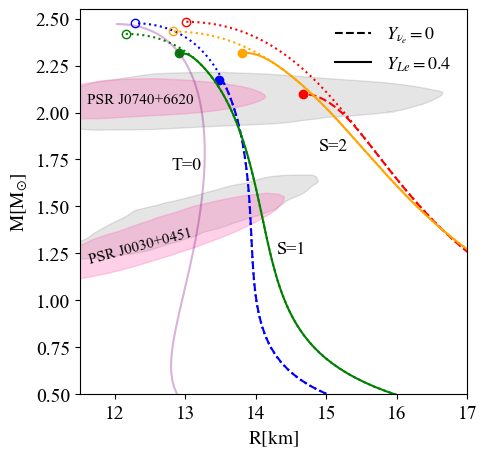}
    \caption{{Gravitational} 
 mass–radius relations for non-rotating
      isentropic stars. {Solid and dashed curves correspond to calculations with and without neutrino trapping, respectively. Results for entropy per baryon
      $S=1$ are shown by the green (solid) and blue (dashed) curves, while those for $S=2$ are shown by the yellow (solid) and red (dashed) curves. The corresponding dotted curves indicate purely hadronic equations of state without a phase transition. For comparison, the mass--radius relation obtained from the cold baryonic equation of state is also shown (purple solid curve). }{The pink and gray shaded ellipses denote the 90\% confidence regions inferred from two independent analyses of}  PSR~{J0030+0451}~
\cite{Riley_2019,Miller_2019}, PSR~{J0740+6620}~\cite{Riley_2021,Miller_2021}. }\label{fig:MR}
\end{figure}

Next, we use the hot and cold equations of state for hadron–quark
matter presented above to compute static, spherically symmetric
configurations of compact stars. The~cold equations of state can be
confronted directly with current astrophysical constraints. The~static
solutions of Einstein’s equations in spherical symmetry, described by
the Tolman–Oppenheimer–Volkoff (TOV)
equations~\cite{Tolman,OppVolkoff}, were obtained using both the cold
EoS and the finite-temperature isentropic EoSs. The~resulting
mass–radius relations are shown in Figure~\ref{fig:MR}.

\vspace{-2pt}

For the cold EoS, we employ the $L_{\rm sym}=50$~MeV and
$Q_{\rm sat}=400$~MeV parametrizations based on the DDME2
framework. The~90\% confidence-level constraints in the mass–radius
diagram are also shown in Figure~\ref{fig:MR}, including the radius
inferences from the NICER observations for canonical
$M\sim1.4\, M_\odot$~\cite{Riley_2019,Miller_2019} and massive $M\sim2\, M_\odot$~\cite{Riley_2021,Miller_2021} neutron stars.  It
can be seen that, in the regime with trapped neutrinos, the maximum masses
(shown by dots) shift toward larger values, while the stellar radii
increase relative to the cold configurations. For~example, for~a star
with $M\sim1.4\, M_\odot$, the radius can be larger by approximately
$1\,\mathrm{km}$ for $S=1$ and a few km for $S=2$ at lepton fraction
$Y_{Le}=0.4$. For~more massive stars $M\sim2\, M_\odot$, these shifts
are somewhat smaller but~still considerable. In~our calculations, the
entropy per baryon and the lepton fraction are assumed to be constant
throughout the star. Although~realistic stellar profiles exhibit
radial variations in these quantities, this approximation provides a
useful first step in exploring the global impact of the temperature and
neutrino trapping on stellar~structures.

These results indicate that thermal effects and neutrino trapping can
significantly modify the global properties of compact stars, including
their maximum masses, radii, and~internal compositions. As~the star
cools and neutrinos gradually escape, its thermodynamic state evolves
toward the cold, neutrino-free configuration. As~illustrated in
Figure~\ref{fig:MR}, this evolution is accompanied by the contraction of
the stellar radius, while the baryonic mass remains approximately
constant. Such a contraction may alter the internal phase structure
and could potentially trigger transitions between different stellar
branches,  most prominently in the case where twin or triplet
configurations are present in the mass--radius diagram~\cite{Alford:2017,Li:2024lmd,Haque2026}.

In closing, we note that classical stable hybrid star sequences
terminate at the maximum mass of the configuration, as~indicated by the dots in Figure~\ref{fig:MR}. Beyond~this point, the~mass decreases as the central density increases, and~the configurations become unstable in the standard turning point-based stability analysis. However, recent studies suggest that the stability properties of hybrid stars may depend sensitively on the dynamical conditions at the interface between hadronic and quark matter. In~particular, if~the phase transition proceeds on a timescale that is slow compared with the oscillation period of the fundamental radial mode, stellar configurations on the descending branch of the mass–radius curve may remain dynamically stable. Such effects arise from the delayed conversion between the two phases and can modify the conventional stability criterion for hybrid stars (see, e.g.,~ref.~\cite{Rau2023,Rau2023b,Celi2025} and references therein). {When applied to binary neutron star mergers, the slow conversion condition requires that the conversion timescale satisfies $\tau_{\mathrm{conv}} \gg \tau_{\mathrm{osc}} \sim 10^{-3} \mathrm{~s}$, where $\tau_{\mathrm{osc}}$ is the post-merger oscillation period, corresponding to frequencies $f_{\rm osc}\sim 1$ kHz. Finite temperatures and neutrino trapping modify the cold dense matter analysis in a twofold way. Firstly, a finite temperature favors faster phase conversion~\cite{Bombaci2011}.  In~this case, the~transition of the quark matter nucleation is expected to be triggered predominantly by thermal nucleation, rather than by quantum tunneling as in cold stars.~On the other hand, as~discussed above, neutrino trapping also shifts the density of the phase transition to higher values. Therefore, neutrino trapping can delay the appearance of quark matter until the matter cools to temperatures where it is transparent to neutrinos, after~which the quark phase can nucleate. If~slow-stable hybrid configurations survive transiently in a hot remnant, they could affect the remnant's characteristics, such as the lifetime, post-merger frequency evolution, etc. Therefore, the~conversion regime is not only a stability problem for static hybrid stars but~also a potentially observable ingredient of the binary neutron star post-merger phenomenology.
}

\section{Discussion}
\label{sec:discussion}

In the present work, we have studied the thermodynamics and stellar
structure of hybrid matter under conditions relevant for
neutrino-trapped proto-neutron stars and binary neutron star merger
remnants. Our results highlight the important roles played by
the temperature, entropy, and~lepton content in determining the onset and
extent of the hadron–quark phase transition, as~well as the global
properties of the resulting hybrid star configurations. In~particular,
we find that neutrino trapping delays the appearance of quark matter
to higher densities and modifies the compositions of both the hadronic
and quark phases. The~mixed phase exhibits a nontrivial thermodynamic
structure, with~a density-dependent pressure that differs from the
behavior obtained in a Maxwell~construction.

Our approach differs in several important respects from earlier
studies. For~example, ref.~\cite{Malfatti2019} employed a nonlocal
extension of the NJL model to describe quark matter, whereas, in this
work, we adopt the local NJL model supplemented by vector interactions
and two-flavor color superconductivity. In~addition, the~phase
transition in ref.~\cite{Malfatti2019} is implemented using a Maxwell
construction at a finite temperature and neutrino chemical potential,
based on the equality of the Gibbs free energy between the two
phases. In~contrast, we construct the transition using a mixed-phase
(Gibbs) prescription that explicitly accounts for baryon and lepton
number conservation and allows us to follow thermodynamic trajectories
at fixed entropy and lepton fractions. While these different treatments
of the phase transition lead to differences in the thermodynamic
structure of the mixed phase, their impact on the global properties of
static stellar configurations is relatively~modest.

Similarly, the~studies in refs.~\cite{Carlomagno2024,Carlomagno_Blaschke} employ a nonlocal
quark model and construct stellar sequences under isothermal
conditions at a fixed lepton chemical potential. Such an approach does
not correspond to configurations with a fixed lepton fraction and
therefore does not directly represent realistic proto-neutron star or
merger remnant conditions. In~contrast, our calculations are performed
for isentropic configurations at a fixed lepton fraction, which provides
a more realistic description of hot, neutrino-trapped compact
stars. Furthermore, the~phase transition in refs.~\cite{Carlomagno2024,Carlomagno_Blaschke} is
also implemented using a Maxwell construction, whereas our treatment
allows for the presence of a mixed phase over a finite density
interval.

{
From an observational standpoint, directly detecting clean signatures of neutrino trapping and finite-temperature effects in binary neutron star mergers is unlikely in the near term (see, however, the~discussion in~\cite{Kyutoku2018}). The~more realistic path is a combined multimessenger analysis: post-merger gravitational waves and kilonova emission interpreted~together.

The post-merger gravitational-wave spectrum is the most promising channel for this. Its characteristic frequencies are set by the remnant's structure, the thermal pressure, and~the trapped neutrino contribution.  Finite-temperature effects shift these frequencies by amounts resolvable with third-generation detectors~\cite{Fields2023}. However, thermal effects on remnant compactness and spectral peaks are strongly correlated with the cold EoS and phase transition parameters, making them difficult to isolate observationally. Progress will therefore require combining the inspiral constraints on the cold EoS along with post-merger observations, which could potentially allow us to disentangle the finite-temperature sector.
}

Overall, the~results presented here demonstrate how the interplay
between temperature, lepton content, and~color-superconducting quark
matter controls the phase structure and stellar properties of hybrid
stars. Future developments will focus on extending the present
framework to three-flavor paired phases and more advanced quark matter
models.

\vspace{6pt}

\acknowledgments{{We} 
 thank M. Alford, D. Blaschke, A. Harutyunyan, M. Oertel and S. Tsiopelas for
discussions.  {This} 
 work was partly supported by the the Polish National
Science Centre (NCN) Grant 2023/51/B/ST9/02798. A. Sedrakian acknowledges  the Deutsche
For\-schungs\-gemeinschaft (DFG)  Grant No. SE 1836/6-1.}


\begin{adjustwidth}{-\extralength}{0cm}
\reftitle{References}


\PublishersNote{}
\end{adjustwidth}

\end{document}